\documentclass[draftcls, onecolumn, 12pt]{IEEEtran}
\usepackage{latexsym,bm}
\usepackage[tbtags]{amsmath}
\usepackage[dvips]{graphicx,color}
\usepackage{subfigure}
\usepackage{amssymb}
\usepackage{cite}
\usepackage{stfloats}

\usepackage{times}
\usepackage{caption}
\usepackage{amsmath}
\usepackage{amssymb}
\usepackage{amscd}
\usepackage{cite}
\usepackage{amsfonts}
\usepackage{graphicx}
\usepackage{psfrag}
\usepackage{epsfig}
\usepackage{algorithm}
\usepackage{algorithmic}
\usepackage{flushend}
\ifCLASSINFOpdf
\else
\fi
\begin{document}
%
\title{Transceiver Design for Cooperative Non-Orthogonal Multiple Access Systems with Wireless Energy Transfer}
\author{ \normalsize Ruijin Sun, Ying Wang$^{*}$, Xinshui Wang  and Yuan Zhang

State Key Laboratory of Networking and Switching Technology, Beijing University of Posts and Telecommunications, Beijing, 100876, China

$^{*}$E-mail: wangying@bupt.edu.cn}

\IEEEaftertitletext{\vspace{-1\baselineskip}}

\maketitle

\begin{abstract}

 In this paper, an energy harvesting (EH) based cooperative non-orthogonal multiple access (NOMA) system is considered, where node S simultaneously sends independent signals to a stronger node R and a weaker node D. We focus on the scenario that the direct link between S and D is too weak to meet the quality of service (QoS) of D. Based on the NOMA principle, node R, the stronger user, has prior knowledge about the information of the weaker user, node D. To satisfy the targeted rate of D, R also serves as an EH decode-and-forward (DF) relay to forward the traffic from S to D. In the sense of equivalent cognitive radio concept, node R viewed as a secondary user assists to boost D's performance, in exchange for receiving its own information from S. Specifically, transmitter beamforming  design, power splitting ratio optimization and receiver filter design to maximize node R's rate are studied with the predefined  QoS constraint of D and the power constraint of S.   Since the problem is non-convex, we propose an iterative approach to solve it. Moreover, to reduce the computational complexity, a  zero-forcing (ZF) based solution is also presented. Simulation results demonstrate that, both two proposed schemes have better  performance than the direction transmission.
\end{abstract}


%
\IEEEpeerreviewmaketitle

\section{Introduction}
%
%
%
%
Self-sustainability and high spectral efficiency are two important metrics for  future wireless communication networks. As a promising solution to enabling self-sustainable communications, radio frequency energy harvesting (RF-EH) technology has recently rekindled considerable interest. The ambient electromagnetic radiation can be captured by the receiver antennas and converted into direct current (DC) voltage \cite{dolgov2010power}. More importantly, RF-EH enables simultaneous wireless information and power transfer (SWIPT) \cite{varshney2008transporting}. To implement it, two practical receiver architectures called time switching (TS) and power splitting (PS) are proposed \cite{zhang2013mimo}. TS switches the receiver between information decoding (ID) and EH modes over time, while PS divides the received signal into two streams with one for ID and the other for EH. For  relay-assisted networks with SWIPT, the energy-constrained relays are allowed to use the harvested RF energy broadcasted by sources to relay the sources' information to destinations. The achievable throughput performance of decode-and-forward (DF) protocol is given in \cite{nasir2014throughput}. Furthermore, relay selection and energy cooperation strategy of multiple users are respectively studied in \cite{men2015joint} and \cite{yang2015energy}. To enhance the EH efficiency, multiple antennas are introduced in relay systems. Joint beamforming design of source and relay node as well as power splitter ratio optimization is investigated in \cite{chen2015joint}. In addition, SWIPT has been extended to cooperative cognitive radio networks \cite{Zheng2014information} and full duplex networks\cite{wang2016transceiver}.

To improve the spectrum efficiency, non-orthogonal multiple access (NOMA) allows multiple users to be served in the same time and frequency resource by using power domain multiplexing. For  user fairness, less powers are allocated to users who have better  channel gains. Moreover, successive interference cancellation (SIC) is adopted by  users with better channel conditions to subtract  signals intended for other users before decoding their own. Based on the power allocation strategy, as proposed in \cite{dingimpact}, NOMA can be classified into two categories, i.e., fixed power allocation NOMA (F-NOMA), and cognitive radio inspired NOMA (CR-NOMA). F-NOMA means that user powers are strictly assigned according to the order of their channel conditions. The performance of downlink NOMA with randomly located user and the impact of user pairing are respectively characterized in \cite{dingimpact} and \cite{ding2014performance} for F-NOMA. Despite that F-NOMA scheme has  superior system performance, it  does not work if multiple antennas are considered. This is owing to the fact that precoders would affect the channel conditions and hence it is challenging to order  users.

As for the CR-NOMA scheme, users with better channel conditions are viewed as secondary users and opportunistically served by the source on the condition that the quality of service (QoS) of weaker users is satisfied. Based on this principle, the analytical outage probability of the stronger user is given in \cite{dingimpact}, since the weaker user's QoS has already been guaranteed. For the multiple-antenna case, a zero-forcing (ZF) based beamforming design and user clustering strategy are investigated for the downlink multiuser NOMA systems \cite{kim2013non}. In that paper, users within the same cluster share the same beamforming vector. To fully exploit the spatial multiplexing gain, two different beamforming vectors respectively for two users are optimized to maximize the system sum rate performance subject to the QoS constraint of the weaker user \cite{sun2015sum}.

It is worth pointing out that the additional introduced secondary users (stronger users) deteriorate the performance of weaker users. In order to improve the reliability of weaker users, cooperative NOMA approach is proposed  \cite{ding2014cooperative}. To be specific, stronger users serve as  relays to forward the traffic from the source to weaker users. It is natural for  stronger users to do this, since the messages intended for weaker users have been decoded and prior known by stronger users if the SIC is successful. In the sense of the equivalent cognitive radio concept,  stronger users would like to relay  messages intended for weaker users, in exchange for receiving their own. This cooperation is especially preferred when direct channels between the source and weaker users are too poor to guarantee their predefined QoS. 

However, the QoS satisfaction  for weaker users is brought by the stronger users' extra transmission power consumption. The  energy shortage at stronger users will break this  cooperation strategy,  even though the channel states between the source and stronger users are well enough for the information cooperation. This motivates us to introduce the wireless energy transfer to  cooperative NOMA systems. That is, the source will transmit both the information and energy to stronger users, in return for  stronger users to boost weaker users' performance. Different from the  user clustering approach and  outage probability given in \cite{liu2015cooperative} with randomly deployed single-antenna users, in this paper, we focus on the beamforming design within one user cluster consisting of two paired users to further enhance the system performance.

 In particular, we consider a RF-EH based cooperative NOMA system in which three nodes are included, i.e., M-antenna  node S, N-antenna  node R and single-antenna  node D.  Node R has a
better connection to node S, while node D, whose service priority is higher, unfortunately has a
worse channel condition. We particularly focus on the the case where the direct link between S and D is too
weak to guarantee the required rate of D. It is a commonly seen situation when the direct link between S and D suffers from a deep fading or the required rate of S is too high. This motivates node
R to simultaneously act as an EH relay to forward the traffic from S to D.
Thus, the cooperative NOMA scheme is proposed. Multiple antennas at relay node are to  enhance the spectral efficiency and energy transfer efficiency.

The main contributions of this work are summarized as follows:

1)  In the proposed three nodes cooperative NOMA system, we focus on the transmitter  beamforming design,  power splitting ratio optimization and the receiver filter design to maximize the rate of R under    constraints that the QoS of D is guaranteed and the transmission power of S is restricted.

2) Due to the coupling nature of variables,  the considered problem is non-convex.  Then, an iterative approach is presented. Specifically, with the fixed receiver filter, the optimal transmitter beamforming and power splitting ratio are obtained via semi-definite relaxation (SDR) and the dual method. With the fixed transmitter beamforming and power splitting ratio, the optimal receiver filter is also derived.

3) Moreover, to reduce the complexity,  ZF-based solution is proposed to find a suboptimal  transmitter beamforming and power splitting ratio with the fixed receiver filter.

4) Comparing these two schemes, the optimal transmitter beamforming scheme always outperforms ZF transmitter
beamforming scheme in terms of node R's rate. Yet, it has almost the same performance with ZF transmitter beamforming scheme in terms of the outage probability of node D. More importantly, both proposed schemes have better outage performance that the direct transmission.

The remainder of the paper is organized as follows.  In Section \uppercase\expandafter{\romannumeral2}, system model and problem formulation are introduced. In Section \uppercase\expandafter{\romannumeral3}, we present an iterative solution to problem $\mathcal{P}$1. In Section \uppercase\expandafter{\romannumeral4}, we further state the ZF-based suboptimal solution to problem $\mathcal{P}2$ to reduce the complexity. The simulation results are presented and discussed in Section \uppercase\expandafter{\romannumeral5}. Finally, Section \uppercase\expandafter{\romannumeral6} concludes the paper.

\emph{Notation}:  Bold lower and upper case letters are used to denote column vectors and matrices, respectively. The superscripts ${{\mathbf{H}}^T}$ and ${{\mathbf{H}}^H}$ is  standard transpose and (Hermitian) conjugate transpose  of $\mathbf{H}$, respectively. $\left\| {\mathbf{h}} \right\|$ refers to the  Euclidean norm of $\mathbf{h}$. $\operatorname{rank}(\mathbf{W})$ and $\operatorname{Tr}(\mathbf{W})$ denote the rank and trace of matrix $\mathbf{W}$, respectively. $\mathbf{W} \succeq \mathbf{0} (\preceq \mathbf0)$ means that matrix $\mathbf{W}$ is positive semidefinite (negative semidefinite). ${\prod _{\mathbf{X}}} = {\mathbf{X}}{\left( {{{\mathbf{X}}^H}{\mathbf{X}}} \right)^{ - 1}}{{\mathbf{X}}^H}$ is the orthogonal projection onto the column space of $\mathbf{X}$, while $\prod _{\mathbf{X}}^ \bot  = {\mathbf{I}} - {\prod _{\mathbf{X}}}$ is the orthogonal projection onto the orthogonal complement of the column space of $\mathbf{X}$.

%

\section{System Model and Problem Formulation}
Considering a cooperative NOMA system,
in which a M-antenna node S simultaneously communicates with a N-antenna node R and a single-antenna node D.
Node R and node D are  users with better and worse connections to S, respectively.
We consider the scenario that  the direct link between S and D is too weak to satisfy the rate demand of the node D.
Therefore, the RF-EH based cooperative NOMA scheme needs to carry out.
In particular, the energy-constrained node R also acts as a relay to first harvest the RF energy broadcasted by S and then uses all the harvested energy to forward the information from S to D.
The PS approach to realize SWIPT is adopted at node R in this paper.  Without loss of generality,
we suppose that $T$ is normalized to be unity. All channels are assumed to be quasi-static,
where the channel coefficients remain the same for each communication duration but vary randomly over different time slots. Note that our considered system model is readily applicable to the downlink transmission with receiver cooperation enhanced 5G systems. In 5G, the
access-point (AP) will serve diverse devices with different capabilities, such as different number
of antennas, different battery capacities, different data requirements, different priorities and so on.


\subsection{Phase 1: Direct Transmission}
During this phase, node S transmits two independent symbols\footnote{On one hand, according to \cite{zhang2013mimo},  single data stream maximizes the harvested energy at EH receiver. So it can substantially benefit the EH-based node R. On the other hand, single-stream provides better diversity gain in terms of the information transmission.} $x_1$ and $x_2$ ($E\left[ {{{\left| {{x_1}} \right|}^2}} \right] = E\left[ {{{\left| {{x_2}} \right|}^2}} \right] = 1$) with power 2$P_s$ to nodes R and D respectively in the same frequency and time slot. The factor 2 is due to the fact that S only transmits signals during the first half duration. The transmitted signal at S can be written as
\begin{equation}
{\bf{x}} = \sqrt {2{P_S}} {{\bf{w}}_1}{x_1} + \sqrt {2{P_S}} {{\bf{w}}_2}{x_2},
\end{equation}
where ${{\bf{w}}_1} \in {\mathbb{C}^{M \times 1}}$ and ${{\bf{w}}_2} \in {\mathbb{C}^{M \times 1}}$ denote the precoding vectors for R and D, respectively. The observations at D and R are respectively given by
\begin{equation}
{y_{D,1}} = \sqrt {2{P_S}} {\bf{h}}_{SD}^H{{\bf{w}}_1}{x_1} + \sqrt {2{P_S}} {\bf{h}}_{SD}^H{{\bf{w}}_2}{x_2} + {n_{D,1}},
\end{equation}
\begin{equation}
{\mathbf{y}_{R,1}} = \sqrt {2{P_S}} {\bf{H}}_{SR}^H{{\bf{w}}_1}{x_1} + \sqrt {2{P_S}} {\bf{H}}_{SR}^H{{\bf{w}}_2}{x_2} + {\bf{n}_{R,1}},
\end{equation}
where ${\bf{h}}_{SD}^{} \in {\mathbb{C}^{M \times 1}}$ and ${\bf{H}}_{SR}^{} \in {\mathbb{C}^{M \times N}}$ denote the channel matrices from S to D and R, respectively. ${n_{D,1}}$ is additive Gaussian white noise (AWGNs) at D  with variances $\sigma _D^2$, and  ${\bf{n}_{R,1}} \in {\mathbb{C}^{N \times 1}}$ is AWGNs vector at R, satisfying ${\bf{ n}_{R,1}} \sim {\cal{CN}}(0, \sigma _R^2{\bf{I}}_N)$.


From (2), the received signal to interference plus noise ratio (SINR) at D to detect $x_2$ is given by
\begin{equation}
{\gamma _{D,1}} = \frac{{2{P_S}{{\left| {{\bf{h}}_{SD}^H{{\bf{w}}_2}} \right|}^2}}}{{2{P_S}{{\left| {{\bf{h}}_{SD}^H{{\bf{w}}_1}} \right|}^2} + \sigma _D^2}}.
\end{equation}

Node R is assumed to be energy-limited and has the ability for RF-EH \cite{nasir2013relaying}. To decode information and harvest energy concurrently, the practical PS-based receiver architecture is applied at node R.  The PS approach works as follows. The node R splits the received RF signal into two streams: one for decoding the information of R and D and the other for harvesting energy to power node R, with the relative power ratio of $\rho$ and $1-\rho$, respectively. The stream flow for information decoding will be converted from the RF to the baseband, and consequently be written as
\begin{equation}
{ {\bf  {y}}^{ID}_{R,1}} = \sqrt \rho  {{\bf{y}}_{R,1}} + {{\bf \tilde{n}}_{R,1}} = \sqrt \rho  \left( {\sqrt {2{P_S}} {\bf{ H}}_{SR}^H{{\bf{w}}_1}{x_1} + \sqrt {2{P_S}} {\bf{ H}}_{SR}^H{{\bf{w}}_2}{x_2} + {{\bf{n}}_{R,1}}} \right) + {{\bf \tilde{n}}_{R,1}},
\end{equation}
where ${{\bf{\tilde{n}}}_{R,1}} \sim {\cal{CN}}(0, \tilde \sigma _R^2 {\bf{I}}_N)$ is the $N\times1$ circuit noise vector caused by the signal frequency conversion from RF to baseband.
After applying  the receiver vector ${{\bf{w}}_R}$, the estimated signal at R can therefore be  represented as
\begin{equation}
{x_{R,1}} ={\bf{w}}_R^H \left[\sqrt\rho  \left( {\sqrt {2{P_S}} {\bf{ H}}_{SR}^H{{\bf{w}}_1}{x_1} + \sqrt {2{P_S}} {\bf{ H}}_{SR}^H{{\bf{w}}_2}{x_2} + {{\bf{n}}_{R,1}}} \right) + {{\bf \tilde{n}}_{R,1}}\right].
\end{equation}

According to the NOMA protocol, SIC is carried out at node R. Specifically, R first decodes the information of D $x_2$ by treating the interference caused by $x_1$ as noise, and then removes this part from the received signal to decode its own information. Mathematically, the received SINRs at R to decode $x_2$ and  $x_1$ can be respectively written as
\begin{equation}
{\gamma _{D,1 \to R,1}} = \frac{{2\rho {P_S}{{\left|{\bf{w}}_R^H {{\bf{ H}}_{SR}^H{{\bf{w}}_2}} \right|}^2}}}{{2\rho {P_S}{{\left| {\bf{w}}_R^H{{\bf{ H}}_{SR}^H{{\bf{w}}_1}} \right|}^2} + \rho \sigma _R^2 \left\| {\mathbf{w}}_R \right\|^2+ \tilde \sigma _R^2\left\| {\mathbf{w}}_R \right\|^2}},
\end{equation}
\begin{equation}
{\gamma _{R,1}} = \frac{{2\rho {P_S}{{\left| {\bf{w}}_R^H{{\bf{ H}}_{SR}^H{{\bf{w}}_1}} \right|}^2}}}{{\rho \sigma _R^2\left\| {\mathbf{w}}_R \right\|^2 + \tilde \sigma _R^2\left\| {\mathbf{w}}_R \right\|^2}},
\end{equation}
which results in the rate of node R ${R_R} = \frac{1}{2}{\log _2}(1 + {\gamma _{R,1}})$.

The signal flow for energy harvesting is
\begin{equation}
{{\bf{y}}^{EH}_{R,1}} = \sqrt {1 - \rho } {{\bf{y}}_{R,1}} = \sqrt {1 - \rho } \left( {\sqrt {2{P_S}} {\bf{ H}}_{SR}^H{{\bf{w}}_1}{x_1} + \sqrt {2{P_S}} {\bf{H}}_{SR}^H{{\bf{w}}_2}{x_2} + {{\bf{n}}_{R,1}}} \right).
\end{equation}

Let $\eta$ denote the energy harvesting efficiency, the harvested energy at R is
\begin{equation}
E = \frac{{\eta (1 - \rho )\left( {2{P_S}\left( {{{\left\| {{\bf{ H}}_{SR}^H{{\bf{w}}_1}} \right\|}^2} + {{\left\| {{\bf{ H}}_{SR}^H{{\bf{w}}_2}} \right\|}^2}} \right)} \right)}}{2}.
\end{equation}
The noise power is ignored compared with the signal power.

We assume that the energy consumed for signal processing is negligible, as compared with the power for signal transmission. Moreover, the transmission period for two phases is equal. Accordingly, the total transmission power at R is
\begin{equation}
{P_R} = 2\eta {P_S}(1 - \rho )\left( {{{\left\| {{\bf{ H}}_{SR}^H{{\bf{w}}_1}} \right\|}^2} + {{\left\| {{\bf{ H}}_{SR}^H{{\bf{w}}_2}} \right\|}^2}} \right).
\end{equation}

\subsection{Phase 2: Cooperative Transmission}

In phase 2,  node S keeps silent, and node R forwards the decoded signal $x_2$ to D with the transmission power $P_R$. That is, DF protocol is used. The received signal at D is
\begin{equation}
{y_{D,2}} = \sqrt {{P_R}} {\bf{h}}_{RD}^H{{\bf{w}}_D}{x_2} + {n_{D,2}},
\end{equation}
where  ${{\bf{h}}_{RD}} \in {{\mathbb{C}}^{N \times 1}}$ and  ${n_{D,2}}\sim {\cal{CN}}(0,\sigma _D^2)$ represent the channel vector from R to D and the AWGN at D, respectively;  ${{\bf{w}}_D}$ is  R's transmit beamforming.
Intuitively, maximal ratio combining (MRC) is the best transmission choice, that is  ${{\bf{w}}_D} = \frac{{{{\bf{h}}_{RD}}}}{{\left\| {{{\bf{h}}_{RD}}} \right\|}} \in {\mathbb{C}^{N \times 1}}$ \cite{tse2005fundamentals},
since only a single data stream is considered here. Then, the received SNR is given by
\begin{equation}
{\gamma _{D,2}} = \frac{{{P_R}{{\left\| {{{\bf{h}}_{RD}}} \right\|}^2}}}{{\sigma _D^2}} = \frac{{2\eta {P_S}(1 - \rho )\left( {{{\left\| {{\bf{ H}}_{SR}^H{{\bf{w}}_1}} \right\|}^2} + {{\left\| {{\bf{ H}}_{SR}^H{{\bf{w}}_2}} \right\|}^2}} \right){{\left\| {{{\bf{h}}_{RD}}} \right\|}^2}}}{{\sigma _D^2}}.
\end{equation}

At the end of this phase, MRC strategy is applied to combine the signal of  ${y_{D,1}}$ and ${y_{D,2}}$. Consequently, the combined SINR at D is
\begin{equation}
\gamma _{D,1,2}^{MRC} = {\gamma _{D,1}} + {\gamma _{D,2}} = \frac{{2{P_S}{{\left| {{\bf{h}}_{SD}^H{{\bf{w}}_2}} \right|}^2}}}{{2{P_S}{{\left| {{\bf{h}}_{SD}^H{{\bf{w}}_1}} \right|}^2} + \sigma _D^2}} + \frac{{2\eta {P_S}(1 - \rho )\left( {{{\left\| {{\bf{ H}}_{SR}^H{{\bf{w}}_1}} \right\|}^2} + {{\left\| {{\bf{ H}}_{SR}^H{{\bf{w}}_2}} \right\|}^2}} \right){{\left\| {{{\bf{h}}_{RD}}} \right\|}^2}}}{{\sigma _D^2}},
\end{equation}
which results in the achievable destination rate ${R_D} = \frac{1}{2}{\log _2}(1 + \gamma _{D,1,2}^{MRC})$.

\subsection{Problem Formulation}
In accordance with the CR-NOMA proposed in \cite{ dingimpact}, the node D, a user with weak channel condition, is viewed as a primary user who occupies the communication channel if orthogonal multiple access (OMA) is used. Based on the equivalent cognitive radio concept, node R is treated as the secondary user to co-work with node D under the underlay mode. Hence, it is of significant importance to meet the predefined QoS of the primary user D, especially when the direct link between S and D cannot satisfy the QoS of D. As a result, in this paper, we aim to maximize the rate of node R subject to the targeted rate constraint of node D and transmission power constraint of S. The optimization problem  can be casted as
\begin{subequations}
\begin{align}
\mathcal{P}1: &~~~~\underset{\;{\mathbf{w}_1},{\mathbf{w}_2},{0\leq\rho \leq1}, \atop
\; \left\|{\bf{w}}_R\right\|^2=1}{\text{max}}&& \frac{{2\rho{P_S} {{\left|{\bf{w}}_R^H {{\mathbf{ H}}_{SR}^H{{\mathbf{w}}_1}} \right|}^2}}}
{{\rho \sigma _R^2 \left\|{\bf{w}}_R\right\|^2+ \tilde \sigma _R^2\left\|{\bf{w}}_R\right\|^2}} \\
&~~~~~~~~\text{s. t.}&&\frac{{2\rho {P_S}{{\left|{\bf{w}}_R^H {{\mathbf{ H}}_{SR}^H{{\mathbf{w}}_2}} \right|}^2}}}
{{2\rho {P_S}{{\left|{\bf{w}}_R^H {{\mathbf{ h}}_{SR}^H{{\mathbf{w}}_1}} \right|}^2} + \rho \sigma _R^2\left\|{\bf{w}}_R\right\|^2 + \tilde \sigma _R^2\left\|{\bf{w}}_R\right\|^2}} \geq{{\gamma}_D'},\displaybreak[0]\\
&&& \!\!\!\!\!\!\!\!\!\!\!\!\!\!\!\!\!\!\!\!\!\!\!\!\!\!\!\!\!\!\!\!\!\!\!\!\!\!\!\!\!\!\!\!\!
\!\!\!\!\!\!\!
\frac{{2{P_S}{{\left| {{\bf{h}}_{SD}^H{{\bf{w}}_2}} \right|}^2}}}{{2{P_S}{{\left| {{\bf{h}}_{SD}^H{{\bf{w}}_1}} \right|}^2} + \sigma _D^2}} + \frac{{2\eta (1 - \rho ){P_S}\left( {{{\left\| {{\bf{ H}}_{SR}^H{{\bf{w}}_1}} \right\|}^2} + {{\left\| {{\bf{ H}}_{SR}^H{{\bf{w}}_2}} \right\|}^2}} \right){{\left\| {{{\bf{h}}_{RD}}} \right\|}^2} }}{{\sigma _D^2}} \ge {{\gamma }_D'},\displaybreak[0]\\
&&&{\left\| {{{\mathbf{w}}_1}} \right\|^2} + {\left\| {{{\mathbf{w}}_2}} \right\|^2} \leq 1,
\end{align}
\end{subequations}
where  ${\gamma '_D} = {2^{2R_D^{\min }}} - 1$ is the minimal SINR threshold at node D with the minimal rate requirement $R_D^{\min }$. What noteworthy is that the constraint (15b) is to ensure that node R can successfully detect node D's information  $x_2$ \cite{ding2014performance}. Different from the single-antenna case where the successful SIC decoding at R is guaranteed by  its better channel gain, beamforming vectors at multiple-antenna S will change the SINRs of R and D. So it becomes necessary to add constraint (15b) \cite{sun2015sum}. Besides, (15c) and (15d) are the rate constraint of D and the transmission power constraint of S, respectively.

\section{Optimization Solution }
In this section, we propose  an iterative approach to solve the non-convex problem $\mathcal{P}1$.

%

\subsection{Step one: Joint optimization of   $\mathbf{w}_1$, $\mathbf{w}_2$ and $\rho$}
With fixed $\mathbf{w}_R$, setting ${\bf{\tilde h}}_{SR}^{} = {\bf{H}}_{SR}^{}{\bf{w}}_R^{} \in {\mathbb{C}^{M \times 1}}$, the problem $\mathcal{P}1$ is simplified as
\begin{subequations}
\begin{align}
\mathcal{P}2: &~~~~\underset{{\mathbf{w}_1},{\mathbf{w}_2},{0\leq\rho \leq1} }{\text{max}}&& \frac{{2\rho{P_S} {{\left| {{\mathbf{ \tilde h}}_{SR}^H{{\mathbf{w}}_1}} \right|}^2}}}
{{\rho \sigma _R^2 + \tilde \sigma _R^2}} \\
&~~~~~~~~\text{s. t.}&&\frac{{2\rho {P_S}{{\left| {{\mathbf{\tilde h}}_{SR}^H{{\mathbf{w}}_2}} \right|}^2}}}
{{2\rho {P_S}{{\left| {{\mathbf{\tilde h}}_{SR}^H{{\mathbf{w}}_1}} \right|}^2} + \rho \sigma _R^2 + \tilde \sigma _R^2}} \geq{{\gamma}_D'},\displaybreak[0]\\
&&& \!\!\!\!\!\!\!\!\!\!\!\!\!\!\!\!\!\!\!\!\!\!\!\!\!\!\!\!\!\!\!\!\!\!\!\!\!\!\!\!\!\!\!\!\!
\!\!\!\!\!\!\!
\frac{{2{P_S}{{\left| {{\bf{h}}_{SD}^H{{\bf{w}}_2}} \right|}^2}}}{{2{P_S}{{\left| {{\bf{h}}_{SD}^H{{\bf{w}}_1}} \right|}^2} + \sigma _D^2}} + \frac{{2\eta (1 - \rho ){P_S}\left( {{{\left\| {{\bf{ H}}_{SR}^H{{\bf{w}}_1}} \right\|}^2} + {{\left\| {{\bf{ H}}_{SR}^H{{\bf{w}}_2}} \right\|}^2}} \right){{\left\| {{{\bf{h}}_{RD}}} \right\|}^2} }}{{\sigma _D^2}} \ge {{\gamma }_D'},\displaybreak[0]\\
&&&{\left\| {{{\mathbf{w}}_1}} \right\|^2} + {\left\| {{{\mathbf{w}}_2}} \right\|^2} \leq 1.
\end{align}
\end{subequations}

Obviously, problem $\mathcal{P}$2 is non-convex, so the key idea to solve it lies in the reformulation of the problem. In order to solve problem $\mathcal{P}$2 efficiently, we introduce a positive variable $\Gamma$ to rewrite the problem as the following $\mathcal{P}$2.1:
\begin{subequations}
\begin{align}
\mathcal{P}2.1: &~~~~\underset{{\mathbf{w}_1},{\mathbf{w}_2},{0\leq\rho \leq1} }{\text{max}}&& \frac{{2\rho{P_S} {{\left| {{\mathbf{\tilde h}}_{SR}^H{{\mathbf{w}}_1}} \right|}^2}}}
{{\rho \sigma _R^2 + \tilde \sigma _R^2}} \\
&~~~~~~~~\text{s. t.}&&\frac{{2\rho {P_S}{{\left| {{\mathbf{\tilde h}}_{SR}^H{{\mathbf{w}}_2}} \right|}^2}}}
{{2\rho {P_S}{{\left| {{\mathbf{\tilde h}}_{SR}^H{{\mathbf{w}}_1}} \right|}^2} + \rho \sigma _R^2 + \tilde \sigma _R^2}} \geq{{\gamma}_D'},\displaybreak[0]\\
&&& \frac{{2{P_S}{{\left| {{\mathbf{h}}_{SD}^H{{\mathbf{w}}_2}} \right|}^2}}}
{{2{P_S}{{\left| {{\mathbf{h}}_{SD}^H{{\mathbf{w}}_1}} \right|}^2} + \sigma _D^2}} \geq \Gamma,\displaybreak[0]\\
&&&\!\!\!\!\!\!\!\!\!\!\!\!\!\!\!\!\!\!\!\!\!\!\!\!\!\!\!\!\frac{{2\eta (1 - \rho ){P_S}\left( {{{\left| {{\mathbf{H}}_{SR}^H{{\mathbf{w}}_1}} \right|}^2} + {{\left| {{\mathbf{H}}_{SR}^H{{\mathbf{w}}_2}} \right|}^2}} \right){{\left\| {{\mathbf{h}_{RD}}} \right\|}^2} }}
{{\sigma _D^2}} \geq {{\gamma}_D'} - \Gamma,\displaybreak[0]\\
&&&{\left\| {{{\mathbf{w}}_1}} \right\|^2} + {\left\| {{{\mathbf{w}}_2}} \right\|^2} \leq 1.
\end{align}
\end{subequations}
Clearly, there exists $\Gamma$ that makes the problem $\mathcal{P}$2.1 identical to problem $\mathcal{P}$2. In the following description, $\Gamma$ is treated as a constant.

We present the optimal solution to problem $\mathcal{P}$2 by applying the celebrated technique of semidefinite relaxation (SDR). Define ${{\mathbf{\tilde H}}_{SR}} = {{\mathbf{\tilde h}}_{SR}}{\mathbf{\tilde h}}_{SR}^H$, ${{\mathbf{\bar H}}_{SR}} = {{\mathbf{H}}_{SR}}{\mathbf{H}}_{SR}^H$, ${{\mathbf{H}}_{SD}} = {{\mathbf{h}}_{SD}}{\mathbf{h}}_{SD}^H$, ${{\mathbf{W}}_1} = {{\mathbf{w}}_1}{\mathbf{w}}_1^H$ and ${{\mathbf{W}}_2}={{\mathbf{w}}_2}{\mathbf{w}}_2^H$ and ignore the rank-one constraint on ${{\mathbf{W}}_1}$ and ${{\mathbf{W}}_2}$, the SDR of problem $\mathcal{P}$2.1 can be expressed as
\begin{subequations}
\begin{align}
\mathcal{P}2.2: &~~~~\underset{{\mathbf{W}_1},{\mathbf{W}_2},{0\leq\rho \leq1} }{\text{max}}&& \frac{{2{P_S}\operatorname{Tr} \left( {{{{\mathbf{\tilde H}}}_{SR}}{{\mathbf{W}}_1}} \right)}}
{{\sigma _R^2 + {{\tilde \sigma _R^2} \mathord{\left/
 {\vphantom {{\tilde \sigma _R^2} \rho }} \right.
 \kern-\nulldelimiterspace} \rho }}} \\
&~~~~~~~~\text{s. t.}&&2{P_S}\operatorname{Tr} \left( {{{{\mathbf{\tilde H}}}_{SR}}{{\mathbf{W}}_2}} \right) \geq {{\gamma}_D'}\left( {2{P_S}\operatorname{Tr} \left( {{{{\mathbf{\tilde H}}}_{SR}}{{\mathbf{W}}_1}} \right) + \sigma _R^2 + {{\tilde \sigma _R^2} \mathord{\left/
 {\vphantom {{\tilde \sigma _R^2} \rho }} \right.
 \kern-\nulldelimiterspace} \rho }} \right),\\
&&& 2{P_S}\operatorname{Tr} \left( {{{\mathbf{H}}_{SD}}{{\mathbf{W}}_2}} \right) \geq \Gamma \left( {2{P_S}\operatorname{Tr} \left( {{{\mathbf{H}}_{SD}}{{\mathbf{W}}_1}} \right) + \sigma _D^2} \right),\\
&&&\operatorname{Tr} \left( {{{{\mathbf{\bar H}}}_{SR}}{{\mathbf{W}}_1}} \right) + \operatorname{Tr} \left( {{{{\mathbf{\bar H}}}_{SR}}{{\mathbf{W}}_2}} \right) \geq\frac{{({{\gamma}_D'} - \Gamma )\sigma _D^2}}
{{2\eta {P_S}{{\left\| {{\mathbf{h}_{RD}}} \right\|}^2}(1 - \rho )}},\\
&&&\operatorname{Tr} \left( {{{\mathbf{W}}_1}} \right) + \operatorname{Tr} \left( {{{\mathbf{W}}_2}} \right) \leq 1.
\end{align}
\end{subequations}

Note that constraints (18b) and (18d) are convex owing to the fact that both $1/\rho$ and $1/(1-\rho)$ are convex functions with respect to $\rho$ with $0<\rho<1$. However, Problem $\mathcal{P}$2.2 is still nonconvex due to its objective function. Fortunately, this objective function is quasi-concave fractional. According to \cite{isheden2012framework}, a positive parameter $t$ can be introduced to formulate a new problem $\mathcal{P}$2.3 which is closely related with $\mathcal{P}$2.2.
\begin{subequations}
\begin{align}
\mathcal{P}2.3: &~~~~\underset{{\mathbf{W}_1},{\mathbf{W}_2},{0\leq\rho \leq1} }{\text{max}}&& 2{P_S}\operatorname{Tr} \left( {{{{\mathbf{\tilde H}}}_{SR}}{{\mathbf{W}}_1}} \right) - t\left( {\sigma _R^2 + {{\tilde \sigma _R^2} \mathord{\left/ {\vphantom {{\tilde \sigma _R^2} \rho }} \right.
 \kern-\nulldelimiterspace} \rho }} \right) \\
&~~~~~~~~\text{s. t.}&&2{P_S}\operatorname{Tr} \left( {{{{\mathbf{\tilde H}}}_{SR}}{{\mathbf{W}}_2}} \right) \geq {{\gamma}_D'}\left( {2{P_S}\operatorname{Tr} \left( {{{{\mathbf{\tilde H}}}_{SR}}{{\mathbf{W}}_1}} \right) + \sigma _R^2 + {{\tilde \sigma _R^2} \mathord{\left/
 {\vphantom {{\tilde \sigma _R^2} \rho }} \right.
 \kern-\nulldelimiterspace} \rho }} \right),\\
&&& 2{P_S}\operatorname{Tr} \left( {{{\mathbf{H}}_{SD}}{{\mathbf{W}}_2}} \right) \geq \Gamma \left( {2{P_S}\operatorname{Tr} \left( {{{\mathbf{H}}_{SD}}{{\mathbf{W}}_1}} \right) + \sigma _D^2} \right),\\
&&&\operatorname{Tr} \left( {{{{\mathbf{\bar H}}}_{SR}}{{\mathbf{W}}_1}} \right) + \operatorname{Tr} \left( {{{{\mathbf{\bar H}}}_{SR}}{{\mathbf{W}}_2}} \right) \geq\frac{a}{1 - \rho },\\
&&&\operatorname{Tr} \left( {{{\mathbf{W}}_1}} \right) + \operatorname{Tr} \left( {{{\mathbf{W}}_2}} \right) \leq 1,
\end{align}
\end{subequations}
where $a=\frac{{({{\gamma}_D'} - \Gamma )\sigma _D^2}}
{2\eta {P_S}{{\left\| {{\mathbf{h}_{RD}}} \right\|}^2}}$. Given $t$ and  $\Gamma$, Problem $\mathcal{P}$2.3 is a convex semidefinite problem (SDP) and can be efficiently solved by off-the-shelf convex optimization solvers, e.g., CVX \cite{Grant2012cvx}.

\textbf{Remark 1}: It is worth pointing out that problem $\mathcal{P}$2.3 belongs to the so-called “separate SDP” \cite{huang2010rank}. Let (${\mathbf{W}}_1^*,{\mathbf{W}}_2^*,{\rho ^*}$) be the optimal solution to problem $\mathcal{P}$2.3. According to [20, Theorem 2.3], the optimal solution to problem $\mathcal{P}$2.3 always satisfies  ${\operatorname{rank} ^2}({\mathbf{W}}_1^*) + {\operatorname{rank} ^2}({\mathbf{W}}_2^*) \leq 4$
, since the number of generalized constraints are 4. We consider the nontrivial case where ${\mathbf{W}}_1^* \ne 0,{\mathbf{W}}_2^* \ne 0$, then ${\text{rank(}}{\mathbf{W}}_1^*{\text{) = }}1$ and ${\text{rank(}}{\mathbf{W}}_2^*{\text{) = }}1$ can be derived. So the SDR problem is tight.

Though the rank-one beamforming vectors can be directly achieved by solving problem $\mathcal{P}$2.3, the computational complexity is high. To  reduce the complexity, we resort to the Lagrangian dual problem of $\mathcal{P}$2.3 for more insightful results.

Since problem $\mathcal{P}$2.3 is convex and satisfies the Slater's condition, its duality is zero. Let $\lambda_1$, $\lambda_2$, $\lambda_3$ and $\lambda_4$ denote the Lagrange multipliers respectively associated with four constraints of problem $\mathcal{P}$2.3. Then, the  Lagrangian function of problem $\mathcal{P}$2.3 is given by
 \begin{equation}
 \begin{split}
  &\mathcal{L}({{\mathbf{W}}_1},{{\mathbf{W}}_{2}}, \rho, {\lambda _1},{\lambda _2},{\lambda _3},{\lambda _4}) = \operatorname{Tr} ({\mathbf{A}}{{\mathbf{W}}_1}) + \operatorname{Tr} ({\mathbf{B}}{{\mathbf{W}}_2}) - \frac{{(t + {\lambda _1}{{\gamma }_D'})\tilde \sigma _R^2}}
{\rho } - \frac{{{\lambda _3}a}}
{{1 - \rho }} \\
& \qquad  \qquad  \qquad  \qquad  \qquad  \qquad \ \ - t\sigma _R^2 - {\lambda _1}{\gamma '_D}\sigma _R^2 - {\lambda _2}\Gamma \sigma _D^2 + {\lambda _4},
\end{split}
\end{equation}
where
\begin{equation}
{\mathbf{A}} = 2{P_S}(1 - {\lambda _1}{\gamma '_D}){{\mathbf{\tilde H}}_{SR}} + {\lambda _3}{{\mathbf{\bar H}}_{SR}}- 2{P_S}{\lambda _2}\Gamma {{\mathbf{H}}_{SD}} - {\lambda _4}{\mathbf{I}},
\end{equation}
\begin{equation}
{\mathbf{B}} = 2{P_S}{\lambda _1}{{\mathbf{\tilde H}}_{SR}} + {\lambda _3}{{\mathbf{\bar H}}_{SR}} + 2{P_S}{\lambda _2}{{\mathbf{H}}_{SD}} - {\lambda _4}{\mathbf{I}}.
\end{equation}
With the Lagrangian function, the dual function of problem $\mathcal{P}$2.3 is expressed as
\begin{equation}
\mathop {\max }\limits_{{{\mathbf{W}}_1} \succeq \mathbf{0},{{\mathbf{W}}_2} \succeq \mathbf{0},0 \leq \rho \leq 1} {\text{  }}\mathcal{L}({{\mathbf{W}}_1},{{\mathbf{W}}_{2}}, \rho, {\lambda _1},{\lambda _2},{\lambda _3},{\lambda _4})
\end{equation}
The optimal dual variables are represented as ($\lambda_1^*, \lambda_2^*, \lambda_3^*, \lambda_4^*$), and hence the optimal $\mathbf{A}$ and $\mathbf{B}$ are denoted as $\mathbf{A}^*$ and $\mathbf{B}^*$, respectively. To guarantee a bounded dual optimal value of (23), $\mathbf{A}^*$ and $\mathbf{B}^*$ must be negative semidefinite. As a result, we can obtain that  $\operatorname{Tr} ({{\mathbf{A}}^*}{\mathbf{W}}_1^*) = 0$ and $\operatorname{Tr} ({{\mathbf{B}}^*}{\mathbf{W}}_2^*) = 0$.
In addition, according to (20) and (23), the optimal power splitter $\rho^*$ must be a solution of the following problem:
\begin{subequations}
\begin{align}
\mathcal{P}2.4: &~~~~~~~\underset{{\rho } }{\text{min}}&& \!\!\!\!\!\!\!\!\!\!\!\!\!\!\!\!\!\!\!\!\!\!\!\!\!\!\!\!\!\!\!\!\!\!\!\!\!\!\!\!\!\!\!
\!\!\!\!\!\!\!\!\!\!\!\!\frac{{(t + {\lambda _1}{{\gamma}_D'})\tilde \sigma _R^2}}
{\rho } + \frac{{{\lambda _3}a}}
{{1 - \rho }}  \\
&~~~~~~~~\text{s. t.}&&\!\!\!\!\!\!\!\!\!\!\!\!\!\!\!\!\!\!\!\!\!\!\!\!\!\!\!\!\!\!\!\!\!\!\!\!\!\!\!\!\!
\!\!\!\!\!\!\!\!\!\!\!\!\!\!0\leq\rho \leq1.
\end{align}
\end{subequations}

\textbf{Proposition 1}: The optimal solution to problem $\mathcal{P}$2.4 is ${\rho ^*} = \frac{b}{{b + \sqrt {bc} }}$ and the optimal value is  $b + c + 2\sqrt {bc}$, where $b =(t + {\lambda _1}{\gamma '_D})\tilde \sigma _R^2 > 0(t > 0), c =a {\lambda _3} > 0$.

Proof: See Appendix A.

\textbf{Proposition 2}: The optimal dual solution $\lambda_3^*$  to problem $\mathcal{P}$2.3 satisfies  $\lambda_3^*>0$.

Proof: See Appendix B.

Define  $\psi ({\lambda _1},{\lambda _2},{\lambda _3},{\lambda _4}) = \mathop {\max }\limits_{{{\mathbf{W}}_1} \succeq  \mathbf{0},{{\mathbf{W}}_2} \succeq  \mathbf{0}, 0 \leq \rho  \leq 1} {\text{  }}\mathcal{L}({{\mathbf{W}}_1},{{\mathbf{W}}_{2}}, \rho ,{\lambda _1},{\lambda _2},{\lambda _3},{\lambda _4})$
, then the Lagrangian dual problem of $\mathcal{P}$2.3 is  ${\text{ }}\mathop {\min }\limits_{{\lambda _1},{\lambda _2},{\lambda _3},{\lambda _4}} \psi ({\lambda _1},{\lambda _2},{\lambda _3},{\lambda _4})$, which is expanded as ($\mathcal{P}$2.5)
\begin{subequations}
\begin{align}
&\underset{{\lambda_1, \lambda_2, \lambda_3, \lambda_4} }{\text{min}} &&\!\!\!\!\!\!\!\!\!\!\!\!\!\!-(t + {\lambda _1}{{\gamma '}_D})\tilde \sigma _R^2 - {\lambda _3}a - 2\sqrt {\tilde \sigma _R^2(t + {\lambda _1}{{\gamma}_D'}){\lambda _3}a}  - t\sigma _R^2 - {\lambda _1}{{\gamma '}_D}\sigma _R^2 - {\lambda _2}\Gamma \sigma _D^2 + {\lambda _4}\\
 &\!\!\!\mathcal{P}2.5:~\text{s. t.}&&{\mathbf{A} }\preceq \mathbf{0}, {\mathbf{B} }\preceq \mathbf{0}, {\lambda _1} \geq 0, {\lambda _2} \geq 0, {\lambda _3} > 0,{\lambda _4} \geq 0.
\end{align}
\end{subequations}
The problem $\mathcal{P}$2.5 is convex, since $\sqrt{{\tilde \sigma _R^2(t + {\lambda _1}{{\gamma}_D'}){\lambda _3}a}
}$ in (25a) is Geometric mean and thus concave \cite{boyd2004convex}. Due to the zero dual gap, problem $\mathcal{P}$2.5 has the same optimal value with problem $\mathcal{P}$2.3.

With the optimal  $\lambda _1^*,\lambda _2^*,\lambda _3^*,\lambda _4^*$
 achieved by problem $\mathcal{P}$2.5, based on \textbf{Proposition 1}, we can obtain   $\rho^*$. Moreover, the complementary slackness condition of problem $\mathcal{P}$2.3 yields to   ${{\mathbf{A}}^*}{\mathbf{W}}_1^* = \mathbf{0}$
and  ${{\mathbf{B}}^*}{\mathbf{W}}_2^* = \mathbf{0}$. Since  ${\text{rank(}}{\mathbf{W}}_1^*{\text{) = }}1$ and ${\text{rank(}}{\mathbf{W}}_2^*{\text{) = }}1$, we have $\operatorname{rank} ({{\mathbf{A}}^*}) = M-1$ and $\operatorname{rank} ({{\mathbf{B}}^*}) = M-1$. Let  $\mathbf{u}_1$ and $\mathbf{u}_2$  be the basis of the null space of  $\mathbf{A}^*$  and  $\mathbf{B}^*$, respectively, and define  ${{\mathbf{W}}_1'} = {{\mathbf{u}}_1}{\mathbf{u}}_1^H$ and ${{\mathbf{W}}_2'} = {{\mathbf{u}}_2}{\mathbf{u}}_2^H$. Since  $\lambda _3^* > 0$, we have
\begin{equation}
\left\{ \begin{aligned}
  &2{P_S}\tau _1^2\operatorname{Tr} \left( {{{{\mathbf{\tilde H}}}_{SR}}{{{\mathbf{W}}}_1'}} \right) - t\left( {\sigma _R^2 + {{\tilde \sigma _R^2} \mathord{\left/
 {\vphantom {{\tilde \sigma _R^2} {{\rho ^*}}}} \right.
 \kern-\nulldelimiterspace} {{\rho ^*}}}} \right) = {d^*},  \\
 &\tau _1^2\operatorname{Tr} \left( {{{{\mathbf{\bar H}}}_{SR}}{{{\mathbf{W}}}_1'}} \right) + \tau _2^2\operatorname{Tr} \left( {{{{\mathbf{\bar H}}}_{SR}}{{{\mathbf{W}}}_2'}} \right) = \frac{a}
{{1 - {\rho ^*}}},  \\
 \end{aligned} \right.
\end{equation}
where $d^*$  is the optimal value of dual problem $\mathcal{P}$2.5 and $\tau_1$, $\tau_2$ are the power allocation coefficients for node R and D, respectively.

Thus, from (26), we have
\begin{equation}
\left\{ \begin{aligned}
  &\tau _1^* = \sqrt {\frac{{{d^*} + t\left( {\sigma _R^2 + {{\tilde \sigma _R^2} \mathord{\left/
 {\vphantom {{\tilde \sigma _R^2} {{\rho ^*}}}} \right.
 \kern-\nulldelimiterspace} {{\rho ^*}}}} \right)}}
{{2{P_S}\operatorname{Tr} \left( {{{{\mathbf{\tilde H}}}_{SR}}{{{\mathbf{W}}}_1'}} \right)}}}, \\
 &\tau _2^* = \sqrt {\frac{{\frac{a}{1-\rho^*} - \tau _1^{*2}\operatorname{Tr} \left(\mathbf{\bar H}_{SR} {{{{\mathbf{W}}}_1'}} \right)}}
{{\operatorname{Tr} \left( \mathbf{\bar H}_{SR}{{{{\mathbf{W}}}_2'}} \right)}}}.
 \end{aligned} \right.
\end{equation}
Then, optimal beamforming vectors are ${\mathbf{w}}_1^* = \tau _1^*{{\mathbf{u}}_1}$ and ${\mathbf{w}}_2^* = \tau _2^*{{\mathbf{u}}_2}$ with given $t$ and $\Gamma$.

\textbf{Remark 2}: Note that $2M$ complex variables and one real variable are to be optimized for problem $\mathcal{P}$2.3, while only four real variables for problem $\mathcal{P}$2.5. Obviously, problem $\mathcal{P}$2.5 has a lower computational complexity than $\mathcal{P}$2.3. Furthermore, the complexity reduction is remarkable as the number of antennas at S grows.

Now, we turn our attention to find the optimal $\Gamma$ and $t$. Given $t$, define the optimal value of problem $\mathcal{P}$2.3 as $\phi (\Gamma )$
 and its dual function as  $ \operatorname{g}
( {{\lambda _1},{\lambda _2},{\lambda _3},{\lambda _4},\Gamma } ) = \mathop {\max }\limits_{{{\mathbf{W}}_1} \succeq \mathbf{0},{{\mathbf{W}}_2} \succeq \mathbf{0},0 \leq \rho  \leq 1} \mathcal{L}({{\mathbf{W}}_1},{{\mathbf{W}}_{2}}, \\\rho ,{\lambda _1},{\lambda _2},{\lambda _3},{\lambda _4},\Gamma )$. Using the zero dual gap, we have  $\phi (\Gamma ) = \mathop {\min }\limits_{{\lambda _1},{\lambda _2},{\lambda _3},{\lambda _4}} \operatorname{g} \left( {{\lambda _1},{\lambda _2},{\lambda _3},{\lambda _4},\Gamma } \right)$. It is easily checked that $\phi (\Gamma )$  is a pointwise minimum of a family of affine function in terms of $\Gamma$ and as a result concave for  $\Gamma$.
So the optimal  $\Gamma^*$  can be found via the one-dimensional search. Based  on (20), the gradient of $\Gamma$ is expressed as
\begin{equation}
\frac{{d\phi (\Gamma )}}
{{d\Gamma }} =  - 2{P_S}\lambda _2^*{\operatorname{Tr}}\left( {{{\mathbf{H}}_{SD}}{\mathbf{W}}_1^*} \right) - \lambda _2^*\sigma _D^2{\text{ + }}\frac{{\lambda _3^{\text{*}}\sigma _D^2}}
{{\eta {{\left\| {{{\mathbf{h}}_{RD}}} \right\|}^2}(1 - {\rho ^*})}}.
\end{equation}

According to the fractional programming \cite{isheden2012framework}, the optimal solution to problem $\mathcal{P}$2.2 is the same with problem $\mathcal{P}$2.3 when
\begin{equation}
F({t^*}) = \mathop {\max }\limits_{{{\mathbf{W}}_1},{{\mathbf{W}}_2},\rho } 2{P_S}\operatorname{Tr} \left( {{{{\mathbf{\tilde H}}}_{SR}}{{\mathbf{W}}_1}} \right) - {t^*}\left( {\sigma _R^2 + {{\tilde \sigma _R^2} \mathord{\left/ {\vphantom {{\tilde \sigma _R^2} \rho }} \right.\kern-\nulldelimiterspace} \rho }} \right) = 0.
\end{equation}
 The optimal $t^*$ can be found by the Dinkelbach method \cite{isheden2012framework}. Therefore, problem $\mathcal{P}$2 is successfully solved. Detailed steps of proposed Algorithm 1 are summarized  as below.

 \begin{algorithm}[h]
    \caption{ The optimal solution to problem $\mathcal{P}$2}
    \begin{algorithmic}[1]
    \STATE {Initialize $t$ satisfying $F(t) \geq 0$ and tolerance $\varepsilon$;}
    \WHILE {$\left( {\left| {F(t )} \right| > \varepsilon } \right)$}
    \STATE {Initialize $\Gamma^{\min}$, $\Gamma^{\max}$ and tolerance $\delta$;}
     \WHILE {$ { \Gamma^{\max}-\Gamma^{\min} > \delta } $}
    \STATE $\Gamma \leftarrow (\Gamma^{\min}+\Gamma^{\max})/2$;
    \STATE Solve problem $\mathcal{P}$2.5 to obtain {$\lambda_1^*$, $\lambda_2^*$,  $\lambda_3^*$, $\lambda_4^*$ and $\rho^*$};
    \STATE Calculate $\mathbf{w}_1^*$ and $\mathbf{w}_2^*$ according to (27) and calculate $\frac{{d\phi (\Gamma )}} {{d\Gamma }}$ according to (28);
    \IF{$\frac{{d\phi (\Gamma )}} {{d\Gamma }} \geq 0$}
    \STATE $\Gamma^{\min} \leftarrow \Gamma $;
    \ELSE
    \STATE $\Gamma^{\max} \leftarrow \Gamma$;
    \ENDIF
    \ENDWHILE
     \STATE $t \leftarrow \frac{2{P_S}{{\left| {{\mathbf{\tilde h}}_{SR}^H{{\mathbf{ w}}_1^*}} \right|}^2}}
            { {\sigma _R^2 + {{\tilde \sigma _R^2} \mathord{\left/ {\vphantom {{\tilde \sigma _R^2} \rho^* }} \right.\kern-\nulldelimiterspace} \rho^* }} }$;
    \ENDWHILE
    \RETURN $\mathbf{ w}_1^*$, $\mathbf{w}_2^*$ and $\rho^*$;
    \label{code:recentEnd}
    \end{algorithmic}
    \end{algorithm}

 \subsection{Step two: Optimization of  $\mathbf{w}_R$} 
With fixed $\mathbf{w}_1$, $\mathbf{w}_2$ and $\rho$, define ${\bf{h}}_{1}^{} = {\bf{H}}_{SR}^{H}{\bf{w}}_1^{} \in {\mathbb{C}^{N \times 1}}$ and ${\bf{h}}_{2}^{} = {\bf{H}}_{SR}^{H}{\bf{w}}_2^{} \in {\mathbb{C}^{N \times 1}}$,
the optimization problem is formulated as
\begin{subequations}
\begin{align}
\mathcal{P}3:&~~~~\underset{{{{\left\| {{{\mathbf{w}}_R}} \right\|}^2} = 1}}{\text{max}}&& \!\!\!\!\!\!\!\!\!\!\!\!\!\!\!\!\!\!\!\!\!\!\!\!\!\!\!\!\!\!
{\left| {{\mathbf{h}}_1^H{{\mathbf{w}}_R}} \right|^2} \\
&~~~~~~~~\text{s. t.}&&\!\!\!\!\!\!\!\!\!\!\!\!\!\!\!\!\!\!\!\!\!\!\!\!\!\!\!\!\!\!
\!\!\!\!\!\!\!\!\!\!\!\!
\!\!\!\!\!\!\!\frac{{2\rho {P_S}{{\left| {{\mathbf{h}}_2^H{{\mathbf{w}}_R}} \right|}^2}}}{{2\rho {P_S}{{\left| {{\mathbf{h}}_1^H{{\mathbf{w}}_R}} \right|}^2} + \rho \sigma _R^2 + \tilde \sigma _R^2}} \geq {{\gamma'_D}}.
\end{align}
\end{subequations}
It is easy to observe that constraint (30b) is active at the optimum. That is,
\begin{equation}
{\text{ }}2\rho {P_S}{\left| {{\mathbf{h}}_2^H{{\mathbf{w}}_R}} \right|^2} = 2\rho {P_S}{\gamma '_D}{\left| {{\mathbf{h}}_1^H{{\mathbf{w}}_R}} \right|^2} + {\gamma '_D}\sigma _R^2 + {\gamma '_D}\tilde \sigma _R^2{\text{  }}.
\end{equation}

Since $\mathbf{w}_R$ is only related to $\mathbf{h}_1$ and $\mathbf{h}_2$, according to \cite{Jor2008complete}, the optimal $\mathbf{w}_R$ can be parametrized as
\begin{equation}
{{\mathbf{w}}_R} = \sqrt \lambda  \frac{{{\prod _{{{\mathbf{h}}_2}}}{{\mathbf{h}}_1}}}
{{\left\| {{\prod _{{{\mathbf{h}}_2}}}{{\mathbf{h}}_1}} \right\|}} + \sqrt {1 - \lambda } \frac{{\prod _{{{\mathbf{h}}_2}}^ \bot {{\mathbf{h}}_1}}}
{{\left\| {\prod _{{{\mathbf{h}}_2}}^ \bot {{\mathbf{h}}_1}} \right\|}}, 0 \leq \lambda \leq 1.
\end{equation}
Then, we have

\begin{equation}
{\tilde f}(\lambda)={\left| {{\mathbf{h}}_1^H{{\mathbf{w}}_R}} \right|^2} = {\left( {\sqrt \lambda  {{\left\| {{\Pi _{{{\mathbf{h}}_2}}}{{\mathbf{h}}_1}} \right\|}} + \sqrt {1 - \lambda } \left\| {\Pi ^ \bot_{{{\mathbf{h}}_2}} {{\mathbf{h}}_1}} \right\|} \right)^2}
\end{equation}
\begin{equation}
\text{and}~~ {\tilde g}(\lambda)= {\left| {{\mathbf{h}}_2^H{{\mathbf{w}}_R}} \right|^2} = \lambda {\left\| {{{\mathbf{h}}_2}} \right\|^2}
\end{equation}
Substituting (33) and (34) into (31), we have
\begin{equation}
{\text{ }}2\rho {P_S}{\tilde g}(\lambda) = 2\rho {P_S}{\gamma '_D}{\tilde f}(\lambda) + {\gamma '_D}\sigma _R^2 + {\gamma '_D}\tilde \sigma _R^2{\text{  }},
\end{equation}
which is a  quadric equation after simple mathematical derivations.

Next, we check that whether (35) has a solution within $[0,1]$.
Set $\alpha={{\left\| {{\Pi _{{{\mathbf{h}}_2}}}{{\mathbf{h}}_1}} \right\|}}$ and $\beta={{\left\| {{\Pi ^\bot_{{{\mathbf{h}}_2}}}{{\mathbf{h}}_1}} \right\|}}$, then ${\tilde f}(\lambda)=\alpha^2\lambda+\beta^2(1-\lambda)+2\alpha \beta \sqrt{\lambda(1-\lambda)}$. So we have ${\tilde f}''(\lambda ) =  - \frac{{\alpha \beta }}
{2}{\lambda ^{ - \frac{3} {2}}}{(1 - \lambda )^{ - \frac{3}{2}}}<0$. Thus, ${\tilde f}(\lambda)$ is a concave function in $\lambda$ with ${\tilde f}(0)=\beta^2$ and ${\tilde f}(1)=\alpha^2$. While ${\tilde g}(\lambda)$ is a linear increasing function with ${\tilde g}(0)=0$ and ${\tilde g}(1)=\left\| \mathbf{h}_2\right\|^2$. Fig. 1 gives a brief relationship between ${g}(\lambda)=2\rho {P_S}\tilde {g}(\lambda)$ and ${f}(\lambda)=2\rho {P_S}{\gamma '_D}\tilde{f}(\lambda) + {\gamma '_D}\sigma _R^2 + {\gamma '_D}\tilde \sigma _R^2$. It is easy to observe that, if and only if ${g}(1)\geq {f}(1)$, i.e.,
\begin{equation}
{\text{ }}2\rho {P_S}\left\| \mathbf{h}_2\right\|^2 \geq 2\rho {P_S}{\gamma '_D}\alpha^2 + {\gamma '_D}\sigma _R^2 + {\gamma '_D}\tilde \sigma _R^2{\text{  }}
\end{equation}
is satisfied, equation (35) has a unique solution within $[0,1]$, i.e.,  problem $\mathcal{P}3$ is feasible.  
Actually, if problem $\mathcal{P}2$ is solvable, problem $\mathcal{P}$3 is feasible. This is because that, at least, the initial point of $\mathbf{w}_R$ is one solution to problem $\mathcal{P}$3. Based on   the roots formula of  the quadric equation,   the optimal $\lambda^*$ can be derived. Therefore, the optimal $\mathbf{w}_R^*$ is obtained.
 \begin{figure}
\centering
\includegraphics[width=9cm, height=6.5cm]{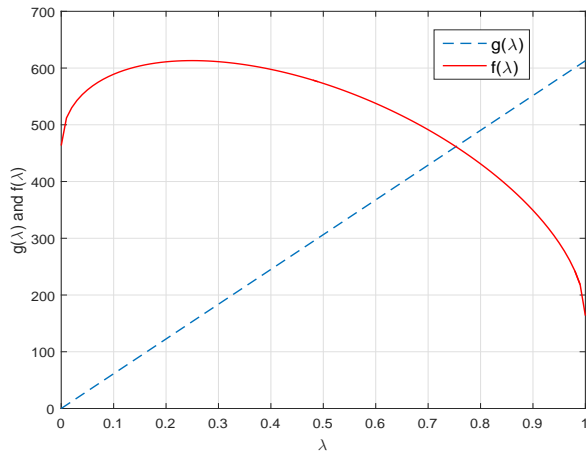}
\caption{A brief relationship between $g(\lambda)$ and $f(\lambda)$.}
\label{systemmodel}
\end{figure}

\subsection{Proposed solution}

 To solve problem $\mathcal{P}$1, we  optimize the transmitter beamforming and power splitter ($\mathbf{w_1}$, $\mathbf{w_2}$, $\rho$)  with the receiver vector ($\mathbf{w_R}$)  iteratively. The approach consists of two steps: (i)  Given  $\mathbf{w_R}$,  optimal (or suboptimal) $\mathbf{w_1^*}$, $\mathbf{w_2^*}$ and $\rho^*$ can be achieved via  Algorithm 1 (or Algorithm 2 listed in Section \uppercase\expandafter{\romannumeral4}); (ii) Given $\mathbf{w_1}$, $\mathbf{w_2}$ and $\rho$,   optimal $\mathbf{w_R^*}$ is obtained by the  solution to problem $\mathcal{P}3$. Repeat these two procedures until problem converges. It is worth pointing out  that the obtained solution to problem $\mathcal{P}1$ can converge. The reason is that, the rate of R increases after each iteration, and the transmission power at S is restricted.  However, since problem $\mathcal{P}1$ is non-convex, convergence to global
maximum is not yet guaranteed.

\section{The ZF-based Suboptimal Solution to problem $\mathcal{P}2$}
Although solving problem $\mathcal{P}$2.5 instead of problem $\mathcal{P}$2.3 can decrease the computational complexity as described in  subsection \emph{$B$} of the previous section, the double round search for finding optimal $\Gamma^*$ and $t^*$ reduces the feasibility of Algorithm 1 to a certain extent in practice. In this section, we  propose a ZF-based suboptimal beamforming scheme to further reduce the complexity of problem $\mathcal{P}$2.

The ZF beamforming is considered to cancel the interference caused by $x_1$ ( the message of node R) at node D. Assume that $\mathbf{w}_1$ lies in the null space of $\mathbf{h}_{SD}^H$, i.e., ${\bf{h}}_{SD}^H{{\bf{w}}_1} = 0$. The singular value decomposition (SVD) of ${\bf{h}}_{SD}^H$ is expressed as ${\bf{h}}_{SD}^H = {\bf{U\Lambda }}{{\bf{V}}^H} = {\bf{U\Lambda }}{[{\bf{\bar V}},{\bf{\tilde V}}]^H}$, where ${\bf{U}} \in {\mathbb{C}^{1 \times 1}}$ and ${\bf{V}} \in {\mathbb{C}^{M \times M}}$ are unitary matrices, ${\bf{\Lambda }} \in {\mathbb{C}^{1 \times M}}$ is a rectangular diagonal matrix. ${\bf{\tilde V}} \in {\mathbb{C}^{M \times (M - 1)}}$ which satisfies ${{\bf{\tilde V}}^H}{\bf{\tilde V}} = {\bf{I}}$ is the last $M-1$ columns of right singular vectors and forms an orthogonal basis for the null space of $\mathbf{h}_{SD}^H$. Thus, ${{\bf{w}}_1} = {\bf{\tilde V}}{{\bf{\tilde w}}_1}$. Problem $\mathcal{P}$2 is consequently formulated as
\begin{subequations}
\begin{align}
\mathcal{P}4: &~~~~\underset{{\mathbf{\tilde w}_1},{\mathbf{w}_2},{0\leq\rho \leq1} }{\text{max}}&& \frac{{2\rho{P_S} {{\left| {{\mathbf{\tilde h}}_{SR}^H{\mathbf{\tilde V}}{{\mathbf{\tilde w}}_1}} \right|}^2}}}
{{\rho \sigma _R^2 + \tilde \sigma _R^2}} \\
&~~~~~~~~\text{s. t.}&&\frac{{2\rho {P_S}{{\left| {{\mathbf{\tilde h}}_{SR}^H{{\mathbf{w}}_2}} \right|}^2}}}
{{2\rho {P_S}{{\left| {{\mathbf{\tilde h}}_{SR}^H{\mathbf{\tilde V}}{{\mathbf{w}}_1}} \right|}^2} + \rho \sigma _R^2 + \tilde \sigma _R^2}} \geq{{\gamma}_D'},\displaybreak[0]\\
&&& \!\!\!\!\!\!\!\!\!\!\!\!\!\!\!\!\!\!\!\!\!\!\!\!\!\!\!\!\!\!\!\!\!\!\!\!\!\!\!\!\!\!\!\!\!
\!\!\!\!\!\!\!
\frac{{2{P_S}{{\left| {{\bf{h}}_{SD}^H{{\bf{w}}_2}} \right|}^2}}}{{ \sigma _D^2}} + \frac{{2\eta (1 - \rho ){P_S}\left( {{{\left| {{\bf{ H}}_{SR}^H{\mathbf{\tilde V}}{{\bf{w}}_1}} \right|}^2} + {{\left| {{\bf{ H}}_{SR}^H{{\bf{w}}_2}} \right|}^2}} \right){{\left\| {{{\bf{h}}_{RD}}} \right\|}^2} }}{{\sigma _D^2}} \ge {{\gamma }_D'},\displaybreak[0]\\
&&&{\left\| {{{\mathbf{\tilde w}}_1}} \right\|^2} + {\left\| {{{\mathbf{w}}_2}} \right\|^2} \leq 1.
\end{align}
\end{subequations}

To effectively tackle problem $\mathcal{P}$4, we slightly reduce its feasible region by multiplying $\frac{{2{P_S}{{\left| {{\bf{h}}_{SD}^H{{\bf{w}}_2}} \right|}^2}}}{{\sigma _D^2}}$, the first term in constraint (16c), by $(1-\rho)$. At the same time, we introduce a positive parameter $t$ to the objective of problem $\mathcal{P}$4, then the reformulated problem $\mathcal{P}$4.1 is written as
\begin{subequations}
\begin{align}
\mathcal{P}4.1: &~~~~\underset{{\mathbf{\tilde w}_1},{\mathbf{w}_2},{0\leq\rho \leq1} }{\text{max}}&& 2{P_S}\rho {\left| {{\bf{\tilde h}}_{SR}^H{\bf{\tilde V}}{{{\bf{\tilde w}}}_1}} \right|^2} - t(\rho \sigma _R^2 + \tilde \sigma _R^2) \\
&~~~~~~~~\text{s. t.}&&\frac{{2\rho {P_S}{{\left| {{\mathbf{\tilde h}}_{SR}^H{{\mathbf{w}}_2}} \right|}^2}}}
{{2\rho {P_S}{{\left| {{\mathbf{\tilde h}}_{SR}^H{\mathbf{\tilde V}}{{\mathbf{w}}_1}} \right|}^2} + \rho \sigma _R^2 + \tilde \sigma _R^2}} \geq{{\gamma}_D'},\displaybreak[0]\\
&&& \!\!\!\!\!\!\!\!\!\!\!\!\!\!\!\!\!\!\!\!\!\!\!\!\!\!\!\!\!\!\!\!\!\!\!\!\!\!\!\!\!\!\!\!\!
\!\!\!\!\!\!\!
\frac{{2(1-\rho){P_S}{{\left| {{\bf{h}}_{SD}^H{{\bf{w}}_2}} \right|}^2}}}{{ \sigma _D^2}} + \frac{{2\eta (1 - \rho ){P_S}\left( {{{\left| {{\bf{ H}}_{SR}^H{\mathbf{\tilde V}}{{\bf{w}}_1}} \right|}^2} + {{\left| {{\bf{H}}_{SR}^H{{\bf{w}}_2}} \right|}^2}} \right){{\left\| {{{\bf{h}}_{RD}}} \right\|}^2} }}{{\sigma _D^2}} \ge {{\gamma }_D'},\displaybreak[0]\\
&&&{\left\| {{{\mathbf{\tilde w}}_1}} \right\|^2} + {\left\| {{{\mathbf{w}}_2}} \right\|^2} \leq 1.
\end{align}
\end{subequations}
According to problem $\mathcal{P}$2.1 and its following reformulations, the SDR of problem $\mathcal{P}$4.1 can be solved by CVX. Obviously, the achieved optimal solution also satisfies the rank-one constraint.

Similarly, we can also tackle problem $\mathcal{P}$4.1 by its Lagrangian dual problem for complexity reduction.  Define $\lambda_1, \lambda_2, \lambda_3$ as dual variables and ${{\bf{\tilde W}}_1} = {{\bf{\tilde w}}_1}{\bf{\tilde w}}_1^H$, ${{\bf{\tilde H'}}_{SR}} = {{\bf{\tilde V}}^H}{\bf{\tilde h}}_{SR}^{}{\bf{\tilde h}}_{SR}^H{\bf{\tilde V}}$, ${{\bf{\bar H'}}_{SR}} = {{\bf{\tilde V}}^H}{\bf{H }}_{SR}^{}{\bf{H}}_{SR}^H{\bf{\tilde V}}$, the  Lagrangian function of problem $\mathcal{P}$4.1 is given by
 \begin{equation}
 \begin{split}
  &\mathcal{L}({{\mathbf{\tilde W}}_1},{{\mathbf{W}}_{2}}, \rho, {\lambda _1},{\lambda _2},{\lambda _3}) = \operatorname{Tr} ({\mathbf{\tilde A}}{{\mathbf{\tilde W}}_1}) + \operatorname{Tr} ({\mathbf{\tilde B}}{{\mathbf{W}}_2}) - \frac{{(t + {\lambda _1}{{\gamma }_D'})\tilde \sigma _R^2}}
{\rho } - \frac{{{\lambda _2}{{\gamma}_D'}\sigma _D^2}}
{{1 - \rho }} \\
& \qquad  \qquad  \qquad  \qquad  \qquad \ \ \ - t\sigma _R^2 - {\lambda _1}{\gamma '_D}\sigma _R^2  + {\lambda _3},
\end{split}
\end{equation}
where ${\bf{\tilde A}} = 2{P_S}(1 - {\lambda _1}{\gamma '_D} ){{\bf{\tilde H'}}_{SR}} +{\lambda _2}\eta {{\bf{\bar H'}}_{SR}}- {\lambda _3}{\bf{I}}$ and ${\bf{\tilde B}} = 2{P_S}{\lambda _1} {{\bf{\tilde H}}_{SR}}+ {\lambda _2}\eta {{\bf{\bar H}}_{SR}} + 2{P_S}{\lambda _2}{{\bf{H}}_{SD}} - {\lambda _3}{\bf{I}}$.
Then, the dual function of problem $\mathcal{P}$4.1 is expressed as
\begin{equation}
\mathop {\max }\limits_{{{{\bf{\tilde W}}}_1} \ge 0,{{\bf{W}}_2} \ge 0,0 \leq \rho  \leq 1} {\rm{  }}\mathcal{L}({{\bf{\tilde W}}_1},{{\bf{W}}_{2}}, \rho ,{\lambda _1},{\lambda _2},{\lambda _3}).
\end{equation}

To ensure that (40) is bounded, $\mathbf{\tilde A}^*$ and $\mathbf{\tilde B}^*$ must be negative semidefinite. As a result, we can obtain that ${\mathop{\rm Tr}\nolimits} ({{\bf{\tilde A}}^*}{\bf{\tilde W}}_1^*) = 0$ and ${\mathop{\rm Tr}\nolimits} ({{\bf{\tilde B}}^*}{\bf{W}}_2^*) = 0$. Similar to \textbf{Proposition 2}, we can prove that $\lambda_2^*>0$. According to \textbf{Proposition 1}, the Lagrangian dual problem of problem $\mathcal{P}$4.1 can be similarly expressed as
\begin{subequations}
\begin{align}
&\underset{{\lambda_1, \lambda_2, \lambda_3 } }{\text{min}} &&\!\!\!\!\!\!\!\!\!\!\!\!\!\!- (t + {\lambda _1}{{\gamma}_D'})\tilde \sigma _R^2 - {\lambda _2}{{\gamma}_D'}\sigma _D^2 - 2\sqrt {\tilde \sigma _R^2(t + {\lambda _1}{{\gamma}_D'}){\lambda _2}{{\gamma}_D'}\sigma _D^2}  - t\sigma _R^2 - {\lambda _1}{{\gamma}_D'}\sigma _R^2 + {\lambda _3}\\
&\!\!\!\mathcal{P}4.2:~\text{s. t.}&&{\mathbf{\tilde A}}  \preceq \mathbf{0}, {\mathbf{\tilde B}}  \preceq \mathbf{0}, {\lambda _1} \geq 0, {\lambda _2} > 0, {\lambda _3} \geq 0,
\end{align}
\end{subequations}
which is convex. Moreover, by using the zero dual gap, we can obtain the optimal solution to problem $\mathcal{P}$4.1. Let ${{\bf{u}}_1}$ and ${{\bf{u}}_2}$ be the basis of the null space of ${{\bf{\tilde A}}^*}$ and ${{\bf{\tilde B}}^*}$, respectively, and define ${{\bf{\tilde W}}_1'} = {{\bf{u}}_1}{\bf{u}}_1^H$ and ${{\bf{W}}_2'} = {{\bf{u}}_2}{\bf{u}}_2^H$. Following the same spirit of (26) and (27), we have
${\bf{\tilde w}}_1^* = \tau _1^*{{\bf{u}}_1}$ and ${\bf{w}}_2^* = \tau _2^*{{\bf{u}}_2}$, where
\begin{equation}
\left\{ \begin{aligned}
  &\tau _1^* = \sqrt {\frac{{{d^*} + t\left( {\sigma _R^2 + {{\tilde \sigma _R^2} \mathord{\left/
 {\vphantom {{\tilde \sigma _R^2} {{\rho ^*}}}} \right.
 \kern-\nulldelimiterspace} {{\rho ^*}}}} \right)}}
{{2{P_S}\operatorname{Tr} \left( {{{{\mathbf{\tilde H}'}}_{SR}}{{{\mathbf{\tilde W}}}_1'}} \right)}}}, \\
 &\tau _2^*=\sqrt {\frac{{\frac{{{{\gamma }_D'}\sigma _D^2}}
{{2{P_S}(1 - \rho^* )}} - \tau_1^{*2}\eta {{\left\| {{{\mathbf{h}}_{RD}}} \right\|}^2}\operatorname{Tr} \left(\mathbf{\tilde H'}_{SR} {{{{\mathbf{W}}}_1'}} \right)}}
{{\operatorname{Tr} \left(\mathbf{ H}_{SD} {{{{\mathbf{W}}}_2'}} \right) + \eta {{\left\| {{{\mathbf{h}}_{RD}}} \right\|}^2}\operatorname{Tr} \left(\mathbf{\bar H'}_{SR} {{{{\mathbf{W}}}_2'}} \right)}}}.
 \end{aligned} \right.
\end{equation}
By adopting Dinkelbach method to search optimal $t^*$, detailed steps of proposed Algorithm 2 are outlined as below.
 \begin{algorithm}[h]
    \caption{ The ZF-based suboptimal solution to problem $\mathcal{P}$4}
    \begin{algorithmic}[1]
    \STATE {Initialize $t$ satisfying $F(t) \geq 0$ and tolerance $\varepsilon$;}
    \WHILE {$\left( {\left| {F(t )} \right| > \varepsilon } \right)$}
    \STATE Solve problem $\mathcal{P}$4.2  to obtain {$\lambda_1^*$, $\lambda_2^*$,  $\lambda_3^*$} and $\rho^*$;
    \STATE Calculate $\mathbf{\tilde w}_1^*$ and $\mathbf{w}_2^*$ according to (42);
    \STATE $t \leftarrow \frac{2{P_S}{{\left| {{\mathbf{\tilde h}}_{SR}^H{\mathbf{\tilde V}}{{\mathbf{\tilde w}}_1^*}} \right|}^2}}
            { {\sigma _R^2 + {{\tilde \sigma _R^2} \mathord{\left/ {\vphantom {{\tilde \sigma _R^2} \rho^* }} \right.\kern-\nulldelimiterspace} \rho^* }} }$;
    \ENDWHILE
    \RETURN $\mathbf{\tilde w}_1^*$, $\mathbf{w}_2^*$ and $\rho^*$;
    \label{code:recentEnd}
    \end{algorithmic}
    \end{algorithm}

\textbf{Remark 3}:
Compared with the optimal beamforming scheme with Algorithm 1, the proposed ZF-based suboptimal beamforming approach with Algorithm 2 further reduces the computational complexity by dropping a round search of $\Gamma$. This reduction is significant since the optimal $\Gamma^*$ needs to be found for every iteration of $t$ in Algorithm 1.

\section{Simulation Results}

In this section, simulation results are presented to evaluate the performance of the proposed schemes. We assume that node S is equipped with $M=2$ antennas, while node R has $N=4$ antennas. We consider a scenario where  channel path losses from node S to R and D are 10 dB and 30 dB, respectively, as well as the path loss from node R to D is 25 dB. The transmission power of S is set to $P_S$ = 30 dB, unless otherwise specified. The variances of noise powers are assumed to unity, i.e., $\sigma _D^2 = \sigma _R^2 = \tilde \sigma _R^2=1$. Moreover, the energy harvesting efficiency is set as 0.8, i.e., $\eta=0.8$. 
Not only the performance of node R's rate, but also the outage probability of node D are evaluated. The optimal scheme and ZF scheme in this section respectively mean the optimal transmitter beamforming scheme and the ZF-based transmitter beamforming scheme. The direct transmission is used as a baseline scheme for the outage performance, which refers to that node S only serves node D with power $P_S$ during the whole time slot. Outage occurs when the required rate of node D cannot be guaranteed. The results in this section are obtained over 500 independent channel realizations, except for Fig. 2.
 \begin{figure}
\centering
\includegraphics[width=9cm, height=6.5cm]{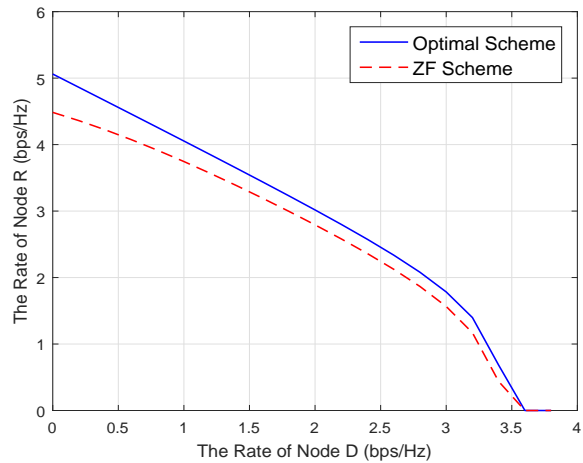}
\caption{Node D - node R rate region for different schemes with $P_S=30$ dB.}
\label{systemmodel}
\end{figure}
 \begin{figure}
\centering
\includegraphics[width=9cm, height=6.5cm]{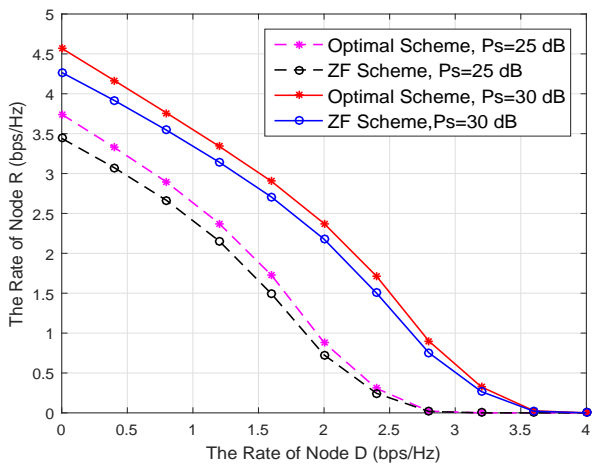}
\caption{Node D - node R average rate region for different schemes with $P_S=30$ dB.}
\label{systemmodel}
\end{figure}

 In Fig. 2,  the rate regions achieved by a specific randomly  chosen channel realization are characterized for different schemes. To be specific, $\mathbf{ H}_{SR}=[0.4035+0.1087i, 0.2944+0.2835i, -0.3285-0.2116i, 0.7751+0.0767i; -0.1413+0.0740i, 0.3469+0.2438i, 0.0396-0.0981i, -0.0480-0.0131i]$, $\mathbf{h}_{SD}=[-0.0137+0.0123i, 0.0054+0.0105i]^T$ and ${\left\| {{{\mathbf{h}}_{RD}}} \right\|^2}=0.0723$. It is observed that the optimal scheme achieves better rate regions than the ZF scheme. In addition, the higher rate node D requires, the smaller gap  the optimal and ZF scheme have.   Then, the impact of transmission power at node S on the achieved rate regions for different schemes are shown in Fig. 3. Observing from this figure, we can see that with  the increasing of  transmission power at node S, the rate regions for both optimal and ZF schemes are greatly enlarged.
\begin{figure}
\centering
\includegraphics[width=9cm, height=6.5cm]{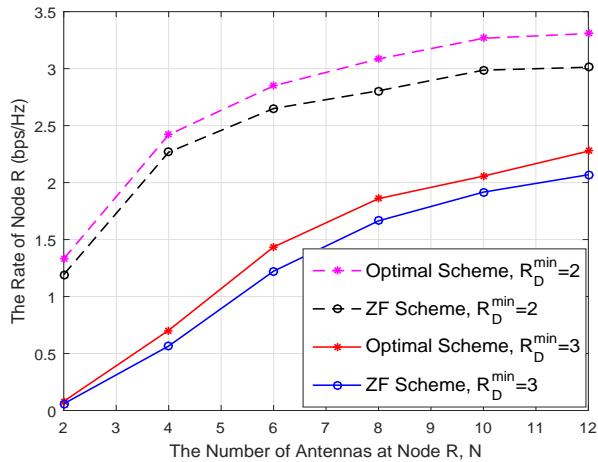}
\caption{The rate of node R versus the number of antennas at node R for different schemes with $P_S=30$ dB.}
\label{systemmodel}
\end{figure}

 Fig. 4 compares the rate of node R for different schemes versus the number of antennas at node R, when  $R_D^{min}$ takes value of 2 bps/Hz and 3 bps/Hz. As excepted, the rate performance of node R is enhanced as the number of antennas grows. Yet the growth trend gradually becomes slow. Besides,  the gap between optimal and ZF schemes in terms of node R's rate is reducing with the increasing of rate requirement of node D.
  \begin{figure}
\centering
\includegraphics[width=9cm, height=6.5cm]{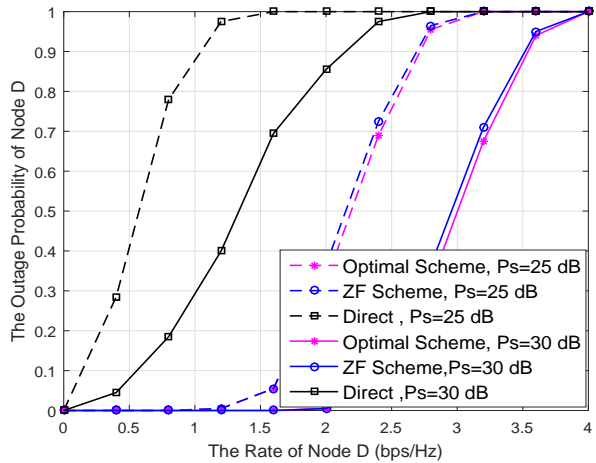}
\caption{The outage probability of node D versus its rate for different schemes.}
\label{systemmodel}
\end{figure}

Next, Fig. 5 presents the outage performance of node D when the rate requirement of D varies from 0 to 4 bps/Hz. It is first noted that the proposed ZF scheme achieves almost the same outage performance with the optimal one. This is owing to the fact that when the rate demand $R_D^{min}$ is extremely close to the outage rate, all  powers should be allocated to beamforming vector $\mathbf{w}_2$ to first satisfy the rate demand of D. So the beamforming  vector $\mathbf{w}_1$  has little effect on the system performance no matter it is designed  optimally or sub-optimally (i.e., ZF-based). This phenomenon is also confirmed in Fig. 2 and Fig. 3 that node R's rate becomes zero almost at the same value of $R_D^{min}$  for the optimal and ZF schemes. More importantly, our proposed two schemes significantly decrease the outage probability of node D compared with the direct transmission. In addition, the higher power  node S transmits, the better outage performance node D has. The outage performance of node D versus the number of antennas at node R  is investigated in Fig. 6. It is observed that as the number of antennas at R increases, the outage probability of node D  reduces obviously.
 \begin{figure}
\centering
\includegraphics[width=9cm, height=6.5cm]{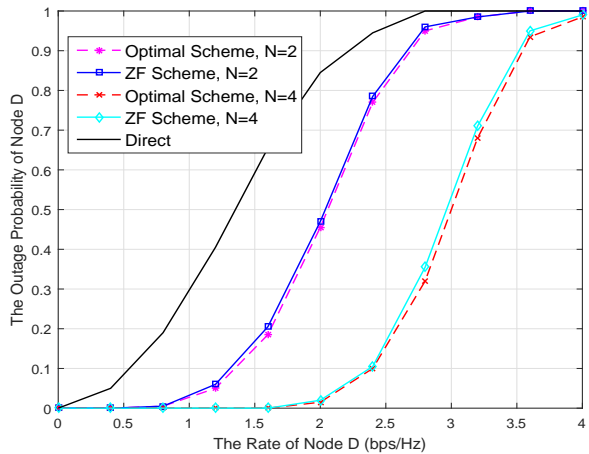}
\caption{The outage probability of node D versus its rate for different schemes different schemes with $P_S=30$ dB.}
\label{systemmodel}
\end{figure}


\section{Conclusion}

In this paper, we have considered an EH-based cooperative NOMA system, where node S simultaneously communicates with a near user, R and a far user, D.  To satisfy the QoS of D, R also serves as an EH DF relay to forward the traffic from S to D.  In particular, transmitter beamforming design, power splitting ratio optimization and receiver filter design to maximize node R's rate have been studied with the predefined QoS constraint of D and the power constraint of S.  Two iterative approaches  have been presented to solve this non-convex problem.  And extensive numerical
experiments have been carried out to evaluate the  performance of our proposed schemes.

\section*{Acknowledgement}
This paper is supported by National Natural Science Foundation
of China (Project 61372113, 61421061).



\appendices
\section{Proof of Proposition 1}
With  ${{ b = }}(t + {\lambda _1}{\gamma '_D})\tilde \sigma _R^2 > 0$ and ${{ c = }}a{\lambda _3} > 0$, problem $\mathcal{P}$2.4 becomes  $\mathop {\min }\limits_{0 \leq \rho  \leq 1} \frac{b}{\rho } + \frac{c}{{1 - \rho }}$. Define   $f{\text{(}}\rho {\text{)  = }}\frac{b}{\rho } + \frac{c}{{1 - \rho }}$. Taking the first derivative of  ${f}(\rho)$ with respective to $\rho$  and setting  $\frac{{df(\rho )}}
{{d\rho }} = 0$. We have  $(c - b){\rho ^2} + 2b\rho  - b = 0$. Then,   $\Delta  = 4{b^2} - 4(c - b)( - b) = 4bc > 0$.

(1) If  $c - b > 0$,  we have ${\rho _1} = \frac{{ - b - \sqrt {bc} }}
{{c - b}} < 0$ and ${\rho _2} = \frac{{ - b + \sqrt {bc} }}{{c - b}} > 0$, where $\rho_1$ and $\rho_2$ are  two roots for $(c - b){\rho ^2} + 2b\rho  - b = 0$. Moreover,  ${\rho _2} = \frac{{ - b + \sqrt {bc} }}
{{c - b}} = \frac{{( - b + \sqrt {bc} )( - b - \sqrt {bc} )}}
{{(c - b)( - b - \sqrt {bc} )}} = \frac{b}
{{b + \sqrt {bc} }} < 1$. So $f(\rho)$  decreases in  $[0, \rho_2]$ and increases in  $[\rho_2, 1]$. Thus,  ${\rho ^*} = {\rho _2} = \frac{b}{{b + \sqrt {bc} }}$.

(2) If  $c-b<0$, we have  ${\rho _1} = \frac{{ - b + \sqrt {bc} }}
{{c - b}} > 0$ and ${\rho _2} = \frac{{ - b - \sqrt {bc} }}{{c - b}} > 0$. In this case, ${\rho _1} = \frac{{ - b + \sqrt {bc} }}
{{c - b}} = \frac{{( - b + \sqrt {bc} )( - b - \sqrt {bc} )}}
{{(c - b)( - c - \sqrt {bc} )}} = \frac{b}{{b + \sqrt {bc} }} < 1$ and ${\rho _2} = \frac{{ - b - \sqrt {bc} }}{{c - b}} = \frac{{( - b - \sqrt {bc} )( - b + \sqrt {bc} )}}{{(c - b)( - b + \sqrt {bc} )}} = \frac{b}{{b - \sqrt {bc} }} > 1$. Thus, $f(\rho)$
decreases in  $[0, \rho_1]$ and increases in  $[\rho_1, 1]$. Therefore,  ${\rho ^*} = {\rho _1} = \frac{b}{{b + \sqrt {bc} }}$.

(3) If  $b=c$,  we have ${\rho ^*} = \frac{1}{2} = \frac{b}{{b + \sqrt {bc} }}$.

Above all,  the optimal  $\rho^*$ is ${\rho ^*} = \frac{b}
{{b + \sqrt {bc} }}$, which results in the optimal value of $\mathcal{P}$2.4  $f({\rho ^*}) = b + c + 2\sqrt {bc}$. This completes the proof.

\section{Proof of Proposition 2}
Observing problem $\mathcal{P}$2.4, note that if  $\lambda_3^*=0$, the optimal solution will be  ${\rho ^*} \to 1$
 (since $t>0$ and then ${(t + {\lambda _1}{{\gamma}_D'})\tilde \sigma _R^2}>0$). As mentioned before, we consider the scenario where the direct link between S and D cannot meet the rate of D. So the required SNR of relay channel is positive, i.e., $\gamma'_D - \Gamma  > 0$, which results in $a>0$ in (19d). So $\rho<1$ must hold. This contradiction indicates that  $\lambda _3^* \ne 0$. Since  $\lambda_3^*\geq 0$, then $\lambda_3^*>0$. This completes the proof.

\ifCLASSOPTIONcaptionsoff
  \newpage
\fi


\begin{thebibliography}{11}   
\footnotesize
 \bibitem{dolgov2010power} Dolgov, A., Zane, R.,  Popovic, Z.: 'Power management system for online low power RF energy harvesting optimization',
IEEE Transactions on Circuits and Systems I: Regular Papers,  2010, \textbf{57},  (7), pp. 1802-1811
 \bibitem{varshney2008transporting} Varshney, L. R.: 'Transporting information and energy simultaneously'.  Proc. IEEE Information Theory (ISIT), 2008,
pp. 1612-1616
\bibitem{zhang2013mimo} Zhang, R. and Ho, C. K.: 'MIMO broadcasting for simultaneous wireless information and power transfer', IEEE Transactions
on Wireless Communications, 2013, \textbf{12},  (5), pp. 1989-2001
\bibitem{nasir2014throughput} Nasir, A. A., Zhou, X., Durrani, S., ~\textit{et al}.: 'Throughput and ergodic capacity of wireless energy harvesting based DF relaying network'. Proc. IEEE International Conference on Communications (ICC), 2014, pp. 4066-4071
\bibitem{men2015joint} Men, J., Ge, J., Zhang, C., ~\textit{et al}.: 'Joint optimal power allocation and relay selection scheme in energy harvesting
asymmetric two-way relaying system', IET Communications, 2015, \textbf{9}, (11), pp. 1421-1426
\bibitem{yang2015energy} Yang, D., Zhou, X., Xiao, L., ~\textit{et al}.:  'Energy cooperation in multi-user wireless-powered relay networks,, IET
Communications, 2015, \textbf{9}, (11), pp. 1412-1420
\bibitem{chen2015joint} Chen, Y., Wen, Z., Beaulieu,N. C., ~\textit{et al}.: 'Joint source-relay design in a MIMO two-hop power-splitting-based
relaying network', IEEE Communications Letters, 2015, \textbf{19}, (10), pp. 1746-1749
 \bibitem{Zheng2014information}Zheng, G., Ho, Z., Jorswieck, E., ~\textit{et al}.: 'Information and Energy Cooperation in Cognitive Radio Networks', IEEE Transactions on Signal Processing, 2014, \textbf{62}, (9), pp. 2290-2303
 \bibitem{wang2016transceiver}Wang,Y., Sun, R., Wang X.: 'Transceiver Design to Maximize the Weighted Sum Secrecy Rate in Full-Duplex SWIPT Systems', IEEE Signal Processing Letters, 2016, \textbf{23}, (6), pp. 883-887
\bibitem{dingimpact} Ding, Z., Fan, P.,  Poor, V.: 'Impact of user pairing on 5G non-orthogonal multiple access downlink transmissions',
IEEE Transactions on Vehicular Technology, 2015, \textbf{PP}, (99)
\bibitem{ding2014performance} Ding, Z., Yang, Z., Fan, P., ~\textit{et al}.: 'On the performance of non-orthogonal multiple access in 5G systems with
randomly deployed users', IEEE Signal Processing Letters, 2014, \textbf{21}, (12), pp. 1501-1505
\bibitem{kim2013non} Kim, B., Lim, S., Kim, H., ~\textit{et al}.:  'Non-orthogonal multiple access in a downlink multiuser beamforming system'.  Proc.
IEEE Military Communications Conference (MILCOM), 2013, pp. 1278-1283
\bibitem{sun2015sum} Sun, Q., Han, S., Xu, Z., ~\textit{et al}.: 'Sum rate optimization for MIMO non-orthogonal multiple
access systems',  Proc. IEEE Wireless Communications and Networking Conference (WCNC), 2015, pp. 747-752
\bibitem{ding2014cooperative} Ding, Z., Peng, M.,  Poor, H. V.: 'Cooperative non-orthogonal multiple access in 5G systems,” IEEE Communications
Letters, 2015, \textbf{19},  (8), pp. 1462-1465
\bibitem{liu2015cooperative} Liu, Y., Ding, Z., Elkashlan, M., ~\textit{et al}.: 'Cooperative non-orthogonal multiple access with simultaneous wireless
information and power transfer', accepted by IEEE Journal on Selected Areas in Communications (JSAC),  2015, online availble: http://120.52.73.77/arxiv.org/pdf/1511.02833.pdf
\bibitem{nasir2013relaying} Nasir, A. A., Zhou, X., Durrani, S., ~\textit{et al}.: 'Relaying protocols for wireless energy harvesting and information
processing', IEEE Transactions on Wireless Communications, 2013,  \textbf{12}, (7), pp. 3622-3636
\bibitem{tse2005fundamentals} Tse, D., and Viswanath, P.: 'Fundamentals of wireless communication' (Cambridge University Press, 2005)
\bibitem{isheden2012framework} Isheden, C., Chong, Z., Jorswieck, E., ~\textit{et al}.: 'Framework for link-level energy efficiency optimization with
informed transmitter', IEEE Transactions on Wireless Communications, 2012, \textbf{11},  (8), pp. 2946-2957
\bibitem{Grant2012cvx} Grant, M.,  Boyd, S.: 'cvx: Matlab software for disciplined convex programming, version 1.22', 2012, online available: http://cvxr.com/cvx.
\bibitem{huang2010rank} Huang, Y. and Palomar, D. P.: 'Rank-constrained separable semidefinite programming with applications to optimal
beamforming', IEEE Transactions on Signal Processing, 2010, \textbf{58}, (2), pp. 664-678
\bibitem{boyd2004convex} Boyd, S. and Vandenberghe, L.: 'Convex optimization' (Cambridge University Press, 2004)
\bibitem{Jor2008complete} Jorswieck, E., Larsson, E., and Danev, D.: 'Complete characterization
of the pareto boundary for the MISO interference channel', IEEE
Transactions on Signal Processing, 2008, \textbf{56},  (10), pp. 5292-5296
\end{thebibliography}
\end{document}